# DNA DENATURATION AS A NEW KIND OF PHASE TRANSITION

Mark Ya. Azbel',

School of Physics and Astronomy, Tel-Aviv University,
Ramat Aviv, 69978 Tel Aviv, Israel[+]
and
Max-Planck-Institute für Festkorperforschung – CNRS,
F38042 Grenoble Cedex 9, France

Abstract

Unbinding of a double-stranded DNA reduces to an unscreened long range interaction and maps on various problems. Heterogeneity renormalizes interaction. Renormalization is temperature dependent. At an unbinding transition it approaches critical dimensionality. This implies giant non-universal critical indexes and invalidity of the Gibbs distribution sufficiently close to the critical temperature $T_c$. Fluctuations are macroscopically large below $T_c$. There are no fluctuations above it.

Thermal unbinding (melting, coiling, denaturation) of a double-stranded DNA molecule is biologically important and physically unique. It yields a phase transition in a one-dimensional system.[1] The system is extraordinary long – the total length of a single mammalian DNA is 1.8m, it consists of ~5 billion nucleotide base pairs. Their sequence is related to genetic information, yet statistically it is close to a random one.[2] The fraction of unbound base pairs as a function of temperature ("the DNA melting curve") is proportional to DNA light absorption at about 260nm. DNA denaturation maps onto a variety of other problems: the binding transition of a polymer onto another polymer, a membrane, or an interface;[3] wetting in two dimensions;[4] depinning of a flux line from a columnar defect in type-II superconductors;[5] localization of a copolymer at a two-fluid interface.[6] DNA denaturation has been extensively studied for nearly four decades.[1-12] Yet, some features of this transition were overlooked. Start with its physics and model. DNA nucleotide base pairs (adenine – thymine AT, guanine – cytosine GC) are large ("mesoscopic") organic molecules. Their unbinding releases few thousand degrees of freedom. The corresponding entropy is $sk_B$ per site[7] ($k_B$ is the Boltzmann constant, s ~ 10). So, while the binding (hydrogen) energy of DNA strands is ~ 3000°K, DNA melts at a relatively low room temperature (~ 300°K), i.e. in the vicinity of the ground state. The Poland-Scheraga model[1] of DNA melting introduces the fusible AT and refractory binding energies $E_1 = -sk_BT_1$ and $E_2 = -sk_BT_2$ correspondingly ($T_1 < T_2$), the boundary energy J per bound segment (J ~ 3000°K accounts for an incomplete unbinding at the boundaries), and the loop entropy $-ck_B \ln L$ per an unbound segment (L is the total number of nucleotide pairs there). The value of the constant c may vary[1,7,10-12] from 1.5 to slightly higher than 2. Thus, at the temperature T, Poland-Scheraga Hamiltonian $E_{\ell Lx}$ of the adjacent bound and melted segments is related to the length $\ell$ and the GC concentration x in the former and to the length L in the latter. Calculated from the energy $-sk_BT$ per site of a completely melted DNA (T is the temperature),

$$E_{\ell Lx} = sk_B \ell \delta T + J + ck_B T \ln L - sk_B \ell (\bar{x} - x) \Delta T.$$

$$\delta T = T - \bar{T}, \quad \bar{T} = T_1 \bar{x} + T_2(1 - \bar{x}), \quad \Delta T = T_2 - T_1,$$

(1)

where $\bar{x}$ is the AT concentration at an entire DNA. Parameters $\bar{T}$ ~ 310K, $\Delta T$ ~ 40K depend on the DNA solution.[7] Start with the well known case of a homopolymer.[1,3,7,8] There $x = \bar{x}$, the last term in Eq. (1) is missing, and $E_{\ell Lx} = E(\ell,L)$ depends on $\ell$ and L only. Then an entire Hamiltonian $H = \sum_n E(\ell^{(n)}, L^{(n)})$, describes an ideal gas of bound and unbound segment pairs $(\ell^{(n)}, L^{(n)})$. It relates the free energy f per site to the normalization condition for the Gibbs probability $p_{\ell L}$ of given $\ell$ and L:

$$p_{\ell L} = \exp\{-(\ell+L)\phi - E(\ell,L)/k_B T\}; \quad \sum_{\ell,L=1}^{\infty} p_{\ell L} = 1, \quad \phi = -f/k_B T \quad (2)$$

When $\phi \ll \exp(-J/k_B T)$, Eqs. (1,2) yield

$$\int_1^{\infty} \exp(-L\phi) L^{-1-c_1} dL = (\phi + \tau) \exp(J/k_B \bar{T}); \quad \tau = s\delta T / \bar{T}; \quad c_1 = c - 1 \quad (3a)$$

Consistent with the Landau-Peierls theorem for the Hamiltonian (1), when $c_1 > 1$, Eq. (3a) yields phase transition. Then, by Eq. (2), $\phi \equiv 0$ does not allow for any excitations of a completely melted polymer. This is specific for the Hamiltonian which depends on $\ell nL$ only – when $L = \infty$, any excitation would imply infinite energy increase. Dependence on $\ell nL$ yields other unusual implications also. Transition is non-universal – its critical indexes depend on $c_1$. Immediately below the critical temperature[1,3,7,8] $T_c$,

$$\phi = \theta \text{ if } c_1 > 1; \quad \phi \sim -\theta \ell n\theta \text{ if } c_1 = 1; \quad \phi \sim \theta^{1/c_1} \text{ if } 1 > c_1 > 0 \tag{3}$$

$$\theta = (T_c - T)/(T_c - \overline{T}); \quad \tau_c = (T_c - \overline{T})/\overline{T} = (1/sc_1)\exp(-J/k_BT);$$

As anticipated, the critical $c_1 = 0$, while $J/k_BT \sim T/\Delta T \sim s \sim 10$ implies, by Eq. (3), a very narrow width of the transition $\sim (T_c - \overline{T})/\overline{T} \sim 10^{-5}$ (i.e. $T_c - \overline{T} \sim 10^{-3} K$), and its very close proximity to the ground state melting temperature $\overline{T}$. Once the free energy (3) is known, the Gibbs probability (2) allows one to calculate any thermodynamic averages and fluctuations. The average (denoted by a bar) relative number $\overline{\omega} = \overline{\ell/L}$ of the melted sites, which is measured via light absorption, the average length $\overline{L}$ of a melted segment, and their relative mean squared fluctuations $\Delta\omega/\overline{\omega}$, $\Delta L/\overline{L}$ are:

$$\overline{\omega} \sim c_1 \exp(J/k_BT) >> 1; \quad \Delta\omega/\overline{\omega} \sim 1$$

$$\overline{L} \sim 1 \text{ if } c_1 > 1; \quad \overline{L} \sim \phi^{c_1 - 1} \text{ if } c_1 < 1 \tag{4a}$$

$$\Delta L/\overline{L} \sim 1 \text{ if } c_1 > 2; \quad \Delta L/\overline{L} \sim \phi^{0.5c_1 - 1} \text{ if } 2 > c_1 > 1; \quad \Delta L/\overline{L} \sim \phi^{-0.5c_1} \text{ if } 1 > c_1 > 0$$

Thus, $\Delta\omega/\overline{\omega}$, $\Delta L/\overline{L}$ are never small, while $\Delta L/\overline{L} \to \infty$ when $T \to T_C$ and $c_1 < 2$. A more physically meaningful fluctuation is

$$\Delta^*\omega/\overline{\omega} \equiv \overline{|\omega - \overline{\omega}|}/\overline{\omega} \sim \overline{\omega}^{-c_1} << 1; \quad \Delta^*L/\overline{L} \equiv \overline{|L - \overline{L}|}/\overline{L} \sim 1 \tag{4b}$$

It demonstrates, in particular, that a characteristic $|L - \overline{L}| \sim \overline{L}$ implies a characteristic $|\ell nL - \ell n\overline{L}| \sim 1$, i.e., $<< \ell n\overline{L}$, when $c_1 < 1$ and $\overline{L} \to \infty$.

Consider heterogeneous DNA. When temperature increases from $\overline{T}$ to $\overline{T} + \delta T$, the Poland-Scheraga Hamiltonian (1) complements the energy increase of an "average" bounded segment (the first three terms) with the energy decrease of a refractory bounded segment (the last term). I prove that in the vicinity of the DNA melting temperature, the last term may be replaced with its thermodynamic average for given lengths of the successive bound and unbound segments. (Such replacement is equivalent to an unusual mean field approximation, which becomes accurate at the phase transition and which technically reduces to a constrained summation in the partition function). The resulting Hamiltonian describes a homopolymer with the renormalized loop entropy. The renormalization, and thus the phase transition singularity it determines, are non-universal and depend on the DNA parameters.

Physics of DNA melting was elucidated in ref. 8. (All following statements are later accurately verified). The segments, rich in the fusible AT, melt first, while the richest in the refractory GC melt last. When c > 1 in Eq. (1), bounded segments completely vanish at a finite critical temperature $T_c$, where L → ∞ and the effective boundary energy $(J+ck_BT\ell nL)$ → ∞. Then the excitation energy also →∞, DNA approaches its ground state, and fluctuations of $\ell$ and x vanish. When T → $T_c$, the length of a ground state (i.e. sufficiently refractory) bounded segment is[8] $\propto \ell nL$ → ∞, to compensate the effective boundary energy in Eq. (1). Sufficiently close to $T_c$, $\ell$ exceeds any finite correlation length, and the probability $w(\ell,x)$ of a given x at such $\ell$ is Gaussian. Since fluctuations of $\ell$ and x vanish at $T_c$, it is ~ the thermodynamic probability $\ell/L$ of a bounded site. So, $\ell$ and x yield $L(\ell,x) \sim \ell/w(\ell,x)$. Thus, the values of $\ell$ and L at a given temperature determine the corresponding value of x according to $w(\ell,x) \sim \ell/L$. The Gaussian w implies $\sqrt{\ell}(\bar{x}-x) \propto \sqrt{\ell n[1/w(\ell,x)]} \gg 1$, where the factor in $w \sim \ell/L$ may be disregarded with negligible error. Thus, $\bar{x} - x$ in Eq. 1 may be replaced with its thermodynamic average according to

$$\ell/L = (\ell/2\pi D^2)^{1/2} \exp(-u^2); \quad u^2 = \ell(\bar{x}-x)^2/2D^2; \quad D^2 = \bar{x}(1-\bar{x}) \qquad (5a)$$

In fact, large $J/k_BT \sim T/\Delta T \sim s \sim 10$ allow for Eq. (5a) already slightly above $\bar{T}$. Equation (1), complemented with the unusual mean field approximation (5a) for x, yield the renormalized Hamiltonian $E^*(\ell,L)$, which depends on the variables $\ell$ and L only. In the leading (in $\ell/L \ll 1$) approximation it equals

$$E^*(\ell,L) = s\ell k_B \delta T + J + ck_B T\ell nL - sk_B D\Delta T\sqrt{2\ell\ell nL} \quad . \qquad (5)$$

The last refractory term accounts for the thermodynamic average of x for given values of $\ell, L, \bar{x}$ and $\Delta T$ in the Poland-Scheraga model for a heteropolymer. The Hamiltonian describes a "renormalized" homopolymer, and Eq. (2), where $E(\ell,L)$ is replaced with $E^*(\ell,L)$, yields its exact free energy. The competition in Eq. (5) of the energy increase and decrease, correspondingly in the "average" first and last "refractory" terms, yields a high and relatively narrow $E^*(\ell,L)$ minimum at the ground state $\ell = \ell_m = 0.5(D\Delta T/\delta T)^2 \ell nL$ (which is indeed $\propto \ell nL$ as stated earlier). The expansion of $E^*(l,L)$ in $l-l_m$ non-universally decreases the factor c in the loop entropy by $s(D\Delta T)^2/2T\delta T$, and Eq. (2), with E replaced with the expanded $E^*$, after a straightforward calculation, yields

$$\int_1^\infty (\ell nL)^{1/2} L^{-1-\delta} \exp(-\phi L) dL = M \qquad (6)$$

where

$$\delta = c_1 - \gamma; \quad \gamma = s(D\Delta T)^2/2\bar{T}\delta T;$$
$$M = \pi^{-1/2}(2c_1)^{-3/2}(sD\Delta T/\bar{T})^2 \exp(J/k_B\bar{T}) \gg 1. \qquad (7)$$

Note that the left hand side of Eq. (6) depends on $\phi$ and $\delta$ only. Thus, Eq. (6) reduces five dimensionless parameters (J/T, $\Delta T/T$, $T/\bar{T}$, $\bar{x}$, c), which determine $\phi$ in a non-renormalized case, to two parameters ($\delta$ and M). When $c_1 < 1$ and $\phi \ll \tau_c-\tau$, Eq. (6) maps onto Eq. (3a), where $c_1$ is renormalized into the temperature dependent $\delta$ (which, unlike $c_1$, may be any sign and which dominates the temperature dependence in the vicinity of the phase transition).

By Eq. (6), $\phi \geq 0$. Since $c_1 > 0$ [1,7,10-12], $\phi = 0$ is achieved (as stated) at finite temperature $T = T_c$. There

$$\delta(T_c) = \delta_c = c_1\theta^*, \quad \theta^* = (2\pi^2)^{1/3}(\overline{T}/sD\Delta T)^{4/3}\exp(-2J/3k_B\overline{T})$$
$$\delta T_c/\overline{T} \simeq (s/2c_1)(D\Delta T/\overline{T})^2, \quad \delta T_c \equiv T_c - \overline{T} \qquad (8)$$

Note that, by Eq. (8), $\delta T_c \sim 3°K$. When $T_c - T \ll \delta T_c$, then $\delta = c_1(\theta^* - \theta)$, where $\theta = (T_c - T)/\delta T_c$ is the relative distance to the critical temperature. By Eq. (4b), $\delta \ll 1$ implies $\delta\ell nL \simeq \delta\ell n\overline{L}$, and thus verifies the derivation of Eq. (5). At $T_c$, by Eq. (8), $\delta_c \sim 0.01$, i.e. it is very close to the critical $c_1 = 0$ – cf Eq. (3). By Eq. (6), $L \propto 1/\phi \to \infty$ when $T \to T_c$. This, and $\ell \simeq \ell_m \propto \ell nL$ verify all previous estimates. When $\delta,\phi \ll 1$, asymptotics in Eq. (6), where $M \gg 1$, yield an unusual non-universal singularity:

$$\phi \sim [(3\theta/4\pi^{1/2}\theta^*)\sqrt{\ell n(\theta^*/\theta)}]^{1/c_1\theta^*}; \quad \text{when } \theta \ll \theta^* \qquad (9a)$$
$$\phi \sim [(2\theta^*/\pi\theta)\sqrt[3]{\ell n(\theta/\theta^*)}]^{3/2c_1\theta}; \quad \text{when } 1 \gg \theta \gg \theta^* \qquad (9b)$$

Consider the implications of Eqs. (7–9b). In natural DNA $J/k_B\overline{T} \sim \overline{T}/\Delta T \sim 10$, $D \sim 1/2$, $c_1 \sim 1$. So, in the immediate vicinity of $T_c$, where $\theta < \theta^* \sim 0.01$, the order of the transition, by Eq. (9a), is $1/c_1\theta^* \sim 100$, i.e. giant. The order is non-universal, it depends on the DNA parameters $T_1, T_2, \overline{x}$. The values of $T_1, T_2$ depend on the ligands and their concentrations in the DNA solutions [7,8], which may be manipulated experimentally. Non-universality in Eqs. (9a, 9b) is related to the competition of the refractory and loop entropy terms in Eq. (5), which renormalizes the loop entropy, and thus the singularity. The width of the transition (9a) is very small, yet macroscopic. The crossover from Eq. (9b) to Eq. (9a) occurs when $(T_c-T)/T_c \sim 10^{-4}$. Then $(T_c - T)/T_c \sim 10^{-4}$, $T_c - T \sim 0.01°K$ (cf $\delta T_c = T_c - \overline{T} \sim 3°K$). In the approximation of Eq. (6), the probability density $P_L$ of a given L is $P_L = M^{-1}(\ell nL)^{1/2}L^{-1-\delta}\exp(-\phi L)$. So, by Eqs. (6, 9b), $\overline{L} \propto 1/\phi \propto \exp[1/(T_c - T)]$ exponentially increases to $\overline{L} \sim 10^{40}$ at the crossover. Thus, even in a solution with $\sim 10^{22}$ DNA nucleotide base pairs, all DNA molecules completely unbind in the interval (9b). So, at a small, yet matroscopic distance $\sim 0.01K$ from $T_c$, the effective long range interaction exceeds any macroscopic size of the system. The system can no more be divided into weakly interacting subsystems, thus the Gibbs distribution is invalid. The fraction of bounded sites is correspondingly small there, and the observably quantity is the temperature of complete melting of a finite DNA. If its length is N, then $\overline{L} = N$ at the temperature $T_N$, when

$$\theta_N = (T_c - T_N)/T_c \sim 1/\ell nN \quad . \qquad (10)$$

The mean fluctuation $\Delta\theta_N$ of $\theta_N$ may be estimated from $\overline{L}(\theta_N + \Delta\theta_N) - \overline{L}(\theta_N) = \Delta^*L(\theta_N)$. Similar to Eq. (4b), $\Delta^*L \sim \overline{L}$, and thus[13]

$$\Delta\theta_N/\theta_N \sim 1/\ell nN \quad . \tag{11}$$

Such fluctuation is macroscopic and easily observable. In DNA this situation is related to mesoscopic size of base pairs (which yields large $J/k_BT \sim s \sim 10$, thus small $\theta^*$), and to DNA heterogeneity. By Eq. (7), heterogeneity effectively replaces fixed $c_1$ with $\delta$. The latter decreases to $\delta_c \ll 1$ (at $T = T_c$) and scales the transition order with $1/\delta_c$ in Eq. (9a) and with $1/\delta$ in Eq. (9b). This may be characteristic of any sufficiently strong long range interaction.

By Eq. (10), natural DNA always yields $\theta \gg \theta^*$, i.e. the essential singularity (9b) in $\theta \propto T_c - T$. (This was predicted in ref. 8). By Eq. (8), it proceeds in the interval $T_c - T \sim 0.01 T_c \sim 3\,°K$. Sufficiently close to $\overline{T}$, the length $\overline{L}$ may reach the correlation length of the sequence. Then the distribution $w(\ell, x)$ becomes non-Gaussian. This alters Eq. (5) and the melting curve $\phi(T)$.

Below $\overline{T}$ DNA is mostly bounded, and only anomalously fusible segments melt. Their probability yields the equation which replaces Eq. (5). Their melting proceeds in an entire interval $\Delta T$. Until sufficiently high temperatures the number of segments, which melt nearly simultaneously, becomes large, the DNA melting curve exhibits their successive melting. It is explicitly seen in experiments.[7,8] Thus, in a general case there are three distinctly different temperature intervals: $\theta^* \sim 0.01$, i.e. $T_c - T \sim 0.03\,°K$; $\theta \sim 1$, i.e. $T_c - T \sim 3K$; and $\Delta T \sim 40\,°K$.

A giant order transition (9a) may be observed only when the total number N of base pairs is much larger than $\overline{L}$ at the crossover to Eq. (9b). This implies $\ell nN > 1/\theta^*$. Since, by Eq. (8), $\theta^* \sim 0.005 D^{-4/3}$, so D must be $< 0.03(\ell nN)^{3/4}$. On the other hand, the derivation of Eq. (6) implied the large renormalized term. At the crossover this means $D > 0.03$. In the interval $0.03 < D < 0.03(\ell nN)^{3/4}$ non-universality of the giant critical index in Eq. (9a) may be studied (e.g., via its dependence on $\Delta T$, which changes together with the concentration of solvents in DNA solution[7]).

Presented theory may be numerically tested. Once the ground state is accurately determined analytically [8], computer simulations allow for the study of its fluctuations.

The approach is applicable to other problems also.

To summarize. DNA unbinding with temperature proceeds from piecewise melting of fusible domains, to essential singularity, to giant ($\sim 1/\theta^* > 100$) order phase transition. The latter may be observed when the AT or GC concentration is between 0.03 and $0.03(\ell nN)^{3/4}$, where N is the total number of nucleotide pairs. In the vicinity of complete melting the Gibbs distribution is invalidated.

Acknowledgement. Financial support from A. von Humboldt award and the J. and R. Meyerhoff chair is appreciated.

[+] Permanent address

## References

1. D. Poland, H. A. Scheraga, J. Chem. Phys. **45**, 1456, 1464 (1966).
2. M. Ya. Azbel', Phys. Rev. Lett. **75**, 168 (1995).
3. M. E. Fisher, J. Stat. Phys. **34**, 667 (1984).
4. G. Forgas, J. M. Luck, Th. M. Nieuwenhuizen, and H. Orland, Phys. Rev. Lett. **57**, 2184 (1986); B. Derrida, V. Hakim, and J. Vannimenus, J. Stat. Phys. **66**, 1189 (1992).
5. D. R. Nelson and V. M. Vinokur, Phys. Rev. **B48**, 13060 (1993).
6. T. Carel, D.A. Huse, L. Leibler, and H. Orland, Europhys. Lett. **8**, 9 (1989).
7. R. M. Wartell and A. S. Benight, Phys. Rep. **126**, 67 (1985); O. Gotoh, Adv. Biophys. **16**, 1 (1983); D. Poland, H. A. Scheraga, *Theory of Helix-Coil Transition in Biopolymers* (Academic, New York, 1970).
8. M. Ya. Azbel', Phys. Rev. Lett. **31**, 589 (1973); Phys. Rev. **A20**, 1671 (1979), M.Ya. Azbel', PNAS USA **76**, 101 (1979); Biopolymers **19**, 61, 81, 95, 1311 (1980).
9. L.-H. Tang and M. Chaté, Phys. Rev. Lett. **86**, 830 (2001).
10. Y. Kafri, D. Mukamel, and L. Peliti, Phys. Rev. Lett. **85**, 4988 (2000).
11. M. E. Fisher, J. Chem. Phys. **45**, 1469 (1966); B. Duplantier, Phys. Rev. Lett. **57**, 941 (1986); J. Stat. Phys. **54**, 581 (1989); L. Schäfer, C. von Ferber, U. Lehr, and B. Duplantier, Nucl. Phys. **B374**, 473 (1992); and refs. therein.
12. E. Carlon, E. Orlandini, and A.L. Stella, Phys. Rev. Lett. **88**, 198101 (2002).
13. If DNA is not closed, then an unbounded segment at a free DNA end does not yield the loop entropy. This should be accounted for at the final stages of unbinding of an open-ended (rather than closed as assumed in the Poland-Scheraga model) DNA.